\documentstyle[nato,numreferences,epsf,graphicx]{crckapb}
\begin{opening}
\title{Anisotropic QCD superfluids}
 
\author{Ji\v r\'{I} Ho\v sek}
\institute{Dept. Theoretical Physics, Nuclear Physics Institute\\
           250 68 \v Re\v z (Prague), Czech Republic}

\end{opening}

\begin{document}

\begin{abstract}
We discuss two instances of anisotropic ordered quantum phases
within QCD at finite baryon densities: (1) Fermionic deconfined 
three-color QCD matter with a spin one quark-quark Cooper pair 
condensate can exhibit distinct quantum i.e. low-temperature ($T$) behaviors 
on macroscopic scales which bona fide can be observed in neutron stars.
(2) Bosonic confined two-color QCD matter with a Bose-Einstein condensate of 
spin-one baryons can exhibit distinct quantum i.e. low-$T$ 
behaviors on macroscopic scales which can bona fide be observed
in numerical lattice experiments.  
\end{abstract}

\section{Introduction}

Understanding QCD starts with understanding its ground state:
At zero baryon density $n_B$ (zero quark chemical potential $\mu$)
and short distances the asymptotic freedom of QCD \cite{AF}
translates into a weakly color-paramagnetic behavior of its perturbative
vacuum \cite{Nielsen-Hughes}. Its excitations are the colored quarks,
and the colored massless gluons. At zero baryon density and large 
distances the structure of QCD vacuum is a nightmare of particle physics.
By expectation its excitations are the colorless hadrons, but their
spectrum is not theoretically understood at present. Be it as it may,
the only hadrons obliged to exist under a decent assumption on the QCD
vacuum are the Nambu-Goldstone pions.

At nonzero quark chemical potential the problem of finding the ground state 
of QCD matter is apparently simpler: Very strong restriction 
on the QCD ground state i.e., its Lorentz invariance, is relaxed. The only 
property which remains sacred is the translation invariance. 
Price for relative simplicity of the ground state is the 
complexity of its excitations. They are of two types: (1) Quasiparticles 
are the excitations carrying the quantum numbers of the quantum
fields defining the system. (2) Collective excitations are effectively described 
by composite operators constructed from the quantum fields.
In both cases the lack of Lorentz invariance implies that the form of 
the dispersion law need not have the Einstein form. Quite often 
it can be reconstructed from thermodynamic behavior of the system.

Above the critical value $n_{B}^{c}$ of $n_B$ which marks the 
confinement-decon\-fi\-ne\-ment phase transition quasiparticles of QCD matter 
are excited by the colored quark, and the colored-gluon quantum fields.  
Consequently, by definition for any number of colors $n_c$ the QCD matter 
at very low $T$ is a many-body system of interacting colored quarks and gluons.
Its ground state is determined by the effective interactions relevant in the 
considered density range. Under rather general assumptions it is 
of the Cooper-pairing type. Non-relativistic condensed-matter relatives are the ordered 
many-fermion systems: the superconductors, the superfluid $^{3}He$ and,
expectedly, the dilute quantum gases of fermionic atoms at extremely low $T$.

For $0 < n_B < n_{B}^{c}$ all excitations of QCD matter are by definition 
collective and colorless. (1) For $n_c$ even (we will discuss explicitly only 
$n_c = 2$) all hadrons are bosons. In particular, the ground-state baryons 
are bosons carrying the baryon number, and spin 0 or 1. They can macroscopically
occupy the ground state in any overall zero-momentum configuration.
Non-relativistic condensed-matter relatives are the superfluid $^{4}He$,
and the dilute gases of bosonic atoms. (2) For $n_c$ odd (we have in mind
only the real world of $n_c = 3$) the ground-state baryons are fermions
carrying spin $1/2$ or $3/2$. Their ground state is most probably of the 
Cooper-pair condensate type.

At present, the ``condensed-matter physics of QCD'' \cite{RW} is quite popular.
Reasons are both intellectual and practical. In this talk we discuss briefly
the low-$T$ properties of two distinct ordered phases of QCD matter characterized
by spontaneous breakdown of rotational symmetry: \\
(1) The two-flavor deconfined 
fermionic quark matter with both spin-0 and spin-1 diquark condensates
is discussed in Sect.2. 
Such a mixture of isotropic and anisotropic color-superconducting phases
could manifest experimentally in the interiors of
the neutron stars.\\
(2) The dilute confined bosonic $n_c = 2$ QCD matter
of one massive flavor with Bose-Einstein condensate of spin-1 baryons
is discussed in Sect.3.  Properties of such matter exhibiting 
eventually superfluidity could be tested 
in numerical lattice QCD experiments. It is interesting though not surprising 
that the phenomenological description of soft Nambu-Goldstone modes
is the same for both cases.  

\section{Anisotropic $n_c=3$ QCD matter of colored quarks}

There are good reasons to expect that the dynamics of the low-$T$
moderately dense deconfined $n_c = 3$ colored-quark matter of two light flavors is 
governed by an effective Lagrangian of the $SU(3)_c \times SU(2)_F \times
U(1)_V \times O(3)$ invariant form (detailed review of the subject with
representative references is Michael Buballa's contribution \cite{Michael}
\begin{equation}
{\cal L}_{eff} = \bar \psi[i \not\!\partial - m +\mu\gamma_{0})\psi + 
{\cal L}_{int}
\nonumber
\end{equation}
where ${\cal L}_{int}$ is a local four-fermion interaction to be determined
experimentally. We make here a commonly accepted assumption that ${\cal L}_{int}$
is attractive in the quark-quark color anti-triplet antisymmetric channel.
According to the Cooper theorem the system spontaneously reorganizes its Fermi sea 
into an energetically more favorable state characterized by the ground-state
BCS-type quark-quark condensates. In accordance with Pauli principle 
their explicit form depends solely upon the details of ${\cal L}_{int}$.

One condensate 
\begin{equation}
\delta \;=\;\langle \psi^T \;C\,\gamma_5\,\tau_2\;\lambda_2\; \psi \rangle~,
\label{delta}
\end{equation}
corresponding in an effective Ginzburg-Landau description to the ground-state
expectation value of a complex spin-0 isospin-0 color anti-triplet order parameter
$\phi$ is ``mandatory''. 

Since in (\ref{delta}) only the quarks of
colors 1 and 2 participate we assume, following the suggestion of the pioneering
Ref.\cite{ARW} that the quark of color 3 undergoes the Cooper pairing in
spin one \cite{aniso}:
\begin{equation}
 \delta' \;=\; \langle \psi^T \;C\,\sigma^{03}\;\tau_2\;\hat P_3^{(c)}\;\psi 
  \rangle~
\label{delta1}
\end{equation}
In an effective Ginzburg-Landau description $\delta^{\prime}$ corresponds to 
the ground-state expectation value of a complex spin-1 order parameter
$\phi^{0n} \equiv \phi_n$, and exhibits spontaneous breakdown of 
the rotational symmetry of the system. 

Fermionic excitations above the condensates (\ref{delta}), (\ref{delta1})
and the chiral-symmetry breaking one $\langle \bar \psi\psi \rangle $ are of
two types:

(i) For colors 1 and 2 the dispersion law of Bogolyubov-Valatin quasiquarks
is isotropic:
\begin{equation}
E_{1}^{\pm}(\vec p) = E_2^{\pm}(\vec p) \equiv E^{\pm}(\vec p) 
= \sqrt{(\epsilon_p\pm\mu)^2 + |\Delta|^2}~.
\label{E1}
\end{equation}
Here $\epsilon_p = \sqrt{\vec p^{\,2} + M^2}$ , $\Delta$ is the energy gap 
determined self-consistently, and found to be of the order $\sim 100$~MeV in model
calculations; $M$ is an effective quark mass related to
$\langle \bar \psi\psi \rangle~$.

(ii) For color 3 the dispersion law is
\begin{equation}
E_3^\pm(\vec p) \;=\; \sqrt{ (\sqrt{M_{\mathit{eff}}^2 + \vec p^{\,2}} \pm \mu_{\mathit{eff}}^2)^2
                            + |\Delta_{\mathit{eff}}'|^2 }~,
\label{E3}
\end{equation}  
where 
$\mu_{\mathit{eff}}^2 = \mu^2 + |\Delta'|^2 \sin^2{\theta}$,
$M_{\mathit{eff}} = M \mu/\mu_{\mathit{eff}}$, and
\begin{equation}
|\Delta_{\mathit{eff}}'|^2 = |\Delta'|^2\,(\cos^2{\theta} + 
\frac{M^2}{\mu_{\mathit{eff}}^2}\,\sin^2{\theta})~. 
\label{Dpeff}
\end{equation}
Here $\cos{\theta} = p_3/|\vec p|$, and the expected spontaneous breakdown of the 
rotational invariance is manifest. It is also worth of writing down explicitly 
the peculiar form of the gap equation for $\Delta^{'}$ 
\begin{equation}
\Delta' = 16H_t\Delta' \int \frac{d^3 p}{(2\pi)^3} \, 
\Big\{\; (1-\frac{{\vec p}_\perp^{\,2}}{s})\frac{1}{E_3^-}
\tanh{\frac{E_3^-}{2T}} 
 + (1+\frac{{\vec p}_\perp^{\,2}}{s})\frac{1}{E_3^+}
\tanh{\frac{E_3^+}{2T}}\Big\}~, 
\label{Deltapgap}
\end{equation}  
where $s = \mu_{\mathit{eff}}(\vec p^{\,2} + M_{\mathit{eff}}^2)^{1/2}$.
It is found by fixing the interaction (the coupling $H_{t}$), calculating the
thermodynamic potential $\Omega(T,\mu)$, and imposing the condition  
$\partial\Omega/\partial{\Delta'}^* = 0$.

Physical consequences of allowing for an anisotropic admixture in 
color superconductor are interesting: First, numerical analysis reveals 
extreme sensitivity of the anisotropic gap $\Delta^{'}$ on the 
chemical potential $\mu$, the details of interaction, and the 
cutoff $\Lambda$ needed to regularize the loop gap-equation integrals.
Numerical values of $\Delta^{'}$ range from O(1 MeV) of early expectations
\cite{ARW} to the ones comparable with $\Delta$. Second, at very 
low temperatures the fermionic specific heat $c_v$ of the system
is dominated by the quasiparticles  of color 3 \cite{aniso}:
\begin{equation}
  c_v \; \approx \; \frac{12}{\pi} \frac{\mu^2 +
    |\Delta'|^2}{|\Delta'|}\,T^2\
 \Big[1 + \frac{\Delta'_0}{T} + \frac{1}{2} 
                 \left(\frac{\Delta'_0}{T}\right)^2
             + \frac{1}{6} \left(\frac{\Delta'_0}{T}\right)^3\Big]\,e^{-\frac{\Delta'_0}{T}} ~,
\label{cvapp}
\end{equation}
It is interesting to notice that the low-lying quasiparticle spectrum
around the minimum
\begin{equation}
    \Delta'_0 = \frac{M |\Delta'|}{\sqrt{\mu^2 + |\Delta'|^2}}~.
\label{M0}
\end{equation}
takes the form
\begin{equation}
  E_3^-(p_\perp, p_3) \approx \sqrt{ \Delta_0^{\prime 2} + v_\perp^2 (p_\perp - p_0)^2
                                  + v_3^2 p_3^2}~,  
\end{equation}
i.e., $\Delta^{\prime}_0$ vanishes for $M=0$, and the specific heat becomes quadratic
in T \cite{leggett}. 

Third, because for $\Delta'$ different from zero the condensate (\ref{delta1})
breaks the $O(3) \times U(1)$ symmetry of the model spontaneously, 
the spectrum of the system should contain the collective
Nambu-Goldstone (NG) modes. Due to the Lorentz-noninvariance of the model
there can be subtleties\cite{NLSS,HoOM98,MiSh01}. The NG spectrum can be analyzed \cite{aniso}
 within an underlying effective Ginzburg-Landau
potential 
\begin{equation}
V(\phi) = -a^2 \phi_n^\dagger\phi_n 
+ \frac{1}{2}\lambda_1(\phi_n^\dagger\phi_n)^2
+ \frac{1}{2}\lambda_2\phi_n^\dagger\phi_n^\dagger\phi_m\phi_m,
\end{equation}
for the complex order parameter $\phi_n$ \cite{HoOM98}, with 
$\lambda_1 + \lambda_2 > 0$ for stability. 
For $\lambda_2 < 0$ the ground state is 
characterized by $\phi_{vac}^{(1)} = (\frac{a^2}{\lambda_1})^{1/2} (0,0,1)$
which corresponds to our Ansatz Eq.~(\ref{delta1}) for the BCS-type
diquark condensate $\delta'$. This solution has the property 
$\langle\vec S\rangle^2 
=(\phi_{vac}^{(1)\dagger} \vec S \phi_{vac}^{(1)})^2 = 0$.
The spectrum of small oscillations above $\phi_{vac}^{(1)}$ consists
of 1+2 NG bosons, all with linear dispersion law: one zero-sound
phonon and two spin waves \cite{HoOM98}. Implying a finite
Landau critical velocity, this fact is crucial for a macroscopic
superfluid behavior of the system \cite{MiSh01}.

\section{Anisotropic $n_c=2$ QCD matter of colorless spin-1 baryons}

For understanding the confining QCD vacuum medium the number of 
colors $n_c$, once bigger than one, does not seem to be a crucial
parameter of the QCD Lagrangian. Introducing the quark chemical potential
into it changes, however, the situation dramatically: According to QCD dogmas
for $n_c$ odd/even the colorless baryons are fermions/bosons. 
Consequently, the quantum i.e. low-T behavior of many-baryon 
systems must be markedly different in worlds with three and two colors.
In particular, and most important, integer-spin many-baryon systems 
can exhibit under specific conditions (low densities, weak, repulsive
two-body interactions) the Bose-Einstein condensation.
This phenomenon certainly implies spontaneous breakdown of the 
global $U(1)$ symmetry generated by the operator of baryon charge, and
in the case of spin-1 baryons also of the $O(3)$ rotational symmetry.
Again, the system should contain phenomenologically important 
and theoretically interesting gapless NG modes. 

The fact that the gedanken world of $n_c=2$ QCD matter at finite $\mu$ is, 
unlike the real $n_c=3$ one, accessible to the first-principle 
lattice computations is alluring: Lattice results can replace  
true experimental data, and provide in principle ideal tests 
of analytic models not only of the very hadron formation but also of 
the hadron-hadron interactions.

At present the majority of $n_c=2$ QCD studies at finite $\mu$ is devoted 
to the regime of approximate chiral symmetry. Assumption of its
spontaneous symmetry breakdown is of course good : (1) It is natural 
because for $n_c=3$ and and the number of flavors $n_F=2,3$ it is an experimental fact. 
(2) It is interesting because (i) dealing with the symplectic
group its pattern is generically different 
from the case of $n_c=3$; (ii) it may yield relativistic vector
condensation \cite{relatvector}.(3) It is predictive because the powerful 
chiral perturbation theory can be employed.

We believe it is both useful and interesting to study the $n_c=2$
low-$T$ integer- spin baryonic matter also in the regime of no chiral symmetry.
We have in mind in particular the simple and conceptually clean case 
of one electrically neutral massive flavor: First, quark masses 
are the parameters external to QCD, and their ratio to $\Lambda_{QCD}$,
another theoretically arbitrary parameter of QCD, can be chosen at will. 
Second, an unknown confining dynamics is not masked by the chiral one. Third, 
according to the QCD dogmas there should exist just one type of the colorless
massive spin-1 baryon described at small density characterized by $\mu$
by a non-relativistic second-quantized Schroedinger field $\phi_{n}$.

Employing with pleasure the principle of the least action we refer again to 
the papers \cite{HoOM98} dealing with the description 
of Bose-Einstein condensation of dilute spin-1 gases: Their analysis 
applies here as it applied also to the formally identical Hamiltonian of
Ginzburg-Landau of Sect.2. The $O(3) \times U(1)$
invariant Hamiltonian density of a weakly-interacting
Bose gas of Bogolyubov  dealing with 
spin-1 baryons $\phi_n$ of mass $m$ is:
\begin{equation}
H = \phi_n^ \dagger(-\frac{\nabla^2}{2m} - \mu)\phi_n +  
\frac{1}{2}\lambda_1(\phi_n^\dagger\phi_n)^2
+ \frac{1}{2}\lambda_2\phi_n^\dagger\phi_n^\dagger\phi_m\phi_m,
\label{bogo}
\end{equation}
In contrast to the many-fermion case of previous section we believe that
the phenomenological baryon-baryon couplings $\lambda_1$ and $\lambda_2$
in (\ref{bogo}) are calculable with reasonable accuracy both 
numerically on the lattice, and within models pretending to describe
the confinement of color. The point is that only one of the two 
generically different patterns of the NG boson spectrum can exist \cite{HoOM98}
for given parameters of the primary QCD Lagrangian:

(i) For $\lambda_1 > 0$, $\lambda_2 > 0$ the homogeneous condensate 
\begin{equation}
\langle \Phi_n \rangle~ = \sqrt{\frac{n_0}{2}}(1,i,0) 
\label{ferro}
\end{equation}
describes a ferromagnetic configuration ($\mu = n_0 \lambda_1$). For very small
momenta the spectrum of small oscillations above the condensate (\ref{ferro})
contains soft NG modes having the dispersion law both linear, and quadratic
in $|\vec p|$.

(ii) For  $\lambda_1 > 0$, $\lambda_2 < 0$ the homogeneous condensate
\begin{equation}
\langle \Phi_n \rangle~ = \sqrt{n_0}(1,0,0) 
\label{polar}
\end{equation}
describes a polar configuration ($\mu = n_0(\lambda_1 + \lambda_2))$. For very small
momenta the spectrum of small oscillations above the condensate (\ref{polar})
contains soft NG modes all having linear dispersion laws.
  
\section{Conclusion}

What we have presented in this contribution are two intuitively simple, 
hopefully interesting, and bona fide testable
illustrations of richness of the condensed matter physics aspects of QCD--
both under reasonable theoretical control.
In particular, lack of Lorentz invariance due to the chemical potential
implies interesting properties of the NG boson spectrum.

What we would have liked to present is less simple: 
We speculate that the finite-$\mu$ studies of 2-color QCD might be 
useful also to condensed-matter physics: Strongly interacting (i.e.\ dense) 
integer-spin baryon system can have a ``superfluid'' ground state without
Bose-Einstein condensate, and hence be relevant for microscopic understanding 
of the superfluid $^4He$. 

This work was supported in part by grant GACR 202/02/0847. I am grateful 
to Ji\v r\'i Adam, Michael Buballa, Micaela Oertel and Adriano Di Giacomo
for many pleasant discussions.

\end{document}